\documentclass [a4paper]{article}

\usepackage{psfig,pstricks,multicol,pst-grad,color,fancybox,graphics}
\usepackage{latexsym}
\usepackage{amsmath}
\usepackage{amssymb}
\usepackage{enumerate}
\usepackage{bbm}

\newcommand{\Bra}[1]{\ensuremath{\langle#1|}}
\newcommand{\Ket}[1]{\ensuremath{|#1\rangle}}
\newcommand{\BraKet}[2]{\ensuremath{\langle #1|#2\rangle}}
\newcommand{\KetBra}[1]{\ensuremath{| #1 \rangle \langle #1 |}}

\newcommand{\Eins}{\ensuremath{\mathbbm 1}}
\newcommand{\HH}{\ensuremath{\mathcal{H}}}
\newcommand{\HS}{\ensuremath{\mathcal{HS}}}
\newcommand{\BE}{\begin{equation}}
\newcommand{\EE}{\end{equation}}
\newcommand{\kommentar}[1]{}
\newcommand{\HSNorm}[2]{\ensuremath{\langle #1|#2\rangle_{HS}}}

\begin{document} 
\title{Experimental detection of entanglement via witness operators
and local measurements} 
\author{O. G\"uhne$^1$, P. Hyllus$^1$, D. Bru\ss$^1$,\\
A.~Ekert$^2$, M. Lewenstein$^1$, C.~Macchiavello$^3$, and A. Sanpera$^1$\\[5mm]
$^1$ Institut f\"ur Theoretische Physik, Universit\"at Hannover,\\
 Appelstra\ss e 2, 30167 Hannover, Germany;\\
$^2$ Department of Applied Mathematics and Theoretical Physics,\\ 
University of Cambridge, Wilberforce Road, \\
Cambridge CB3 0WA, UK;\\
$^3$ Dipartimento di Fisica ``A. Volta" and INFM-Unit\'a 
di Pavia,\\
Via Bassi 6, 27100 Pavia, Italy
}
\maketitle    

\noindent
{\bf Abstract:} 
In this paper we address the problem of detection of 
entanglement using only few local measurements when some knowledge
about the state is given. The idea is based on an optimized 
decomposition of witness operators into local operators. 
We discuss two possible ways of optimizing this local decomposition. 
We present several analytical results and estimates for optimized 
detection strategies for NPT states of $2 \times 2$ and $N \times M$ 
systems, entangled states in 3 qubit systems, and bound entangled 
states in $3\times3$ and $2\times4$ systems.

\section{Introduction}
One of the key problems in experiments in quantum information theory 
is the generation and detection of entanglement \cite{gener}. The 
generation of 
entangled states is practically always aimed at particular 
states that can then be used for various applications in 
quantum information processing. Very often one wants to produce certain 
pure states. The generation process in a laboratory is, however, never 
free of imperfections and noise. The states that one produces may, 
but not necessarily have the desired properties. In particular they may, 
but do not have to be entangled. For this reason it is important to develop 
efficient and easy to apply experimental procedures to detect entanglement. 

Obviously, the ultimate goal of detection is to characterize the entanglement
quantitatively, and to identify the regions in the parameter space
which would allow to maximize entanglement 
useful for the particular application.  Before this ambitious goal 
is realized, however, it is important to know whether the state 
one is dealing with is
entangled at all, or not. The reasons for that are at least threefold:
i) it is interesting from the fundamental point of view; ii) it is 
important to know it before applying more complex experimental tools 
to quantify entanglement; iii)  it is essential, if one wants to use the 
generated states for any of the distillation or purification 
protocols \cite{dist}.

For the detection of 
entanglement several strategies are known: First of all
there is the possibility 
of quantum state tomography \cite{tomo}, and then direct application of 
known necessary or sufficient entanglement criteria \cite{primer}. 
Determining the
density matrix via quantum state tomography requires, however,  typically 
a lot of measurements. Note also 
that only for $2\times2$ and $2\times3$ systems necessary and sufficient 
criterion is 
known, the famous Peres-Horodecki criterion of positivity of partial 
transpose \cite{perescrit}. In general, it might be difficult to 
find a sufficient criterion for a given state. 
One could also look for a 
violation of some Bell inequalities, although there exist entangled 
states (bound entangled states)
that do not violate any known Bell-like inequality \cite{bell}.
In fact, there is even a 
conjecture that these states admit a local hidden variable model 
\cite{peres1}.
Looking for a violation of Bell inequalities may thus not be sufficient.

 Recently, there have been several
proposals for the detection of entanglement without estimating the whole 
density matrix \cite{huelga,horodecki1,horodeckiekert1}.  
Although being attractive from both experimental and theoretical 
point of view,  these possibilities have 
some disadvantages. 
The recent proposals require 
collective measurements on several qubits or the construction 
of quantum gates and networks. These requirements are possible, but not easy 
to fulfill with the present experimental techniques of local measurements. 

On the other hand, the methods for the detection of entanglement
mentioned above are in some sense too general for experiments. 
They assume that no {\it a priori} knowledge about the density 
matrix is given. But, as we already mentioned, 
in an experimental situation one usually tries 
to prepare some particular  state. Although one fights with the problems of 
noise and imperfections, in any case the produced state 
cannot be  considered
completely arbitrary.

In this paper, which expands our earlier study \cite{wir},
we address the problem of checking whether a state 
$\varrho$ is entangled or not, when some knowledge about the density 
matrix is given. We want to solve this problem using only local 
von Neumann measurements which can be implemented in a laboratory 
using the present techniques. We also aim at using the smallest
possible number of measurements.

The scheme we use for the detection of entanglement relies on the
well known concept of witness operators \cite{witnesses1, witnesses2}. 
Let us briefly recall what these operators are. A Hermitean operator $W$ 
is called an entanglement witness detecting the entangled state 
$\varrho_e$ if $Tr(W\varrho_e)<0$ and $Tr(W\varrho_s)\geq 0$  for 
all separable states $\varrho_s.$ So, if we have a state $\varrho$ 
and we measure $Tr(W\varrho)<0,$ we can be sure that $\varrho$ is 
entangled. For every entangled state there exists an entanglement 
witness. There has also been an enormous progress in constructing 
witnesses for different classes of states \cite{witnesses2}. 

After  having constructed a witness operator we decompose it into
a weighted sum of projectors onto product vectors. In this way the 
expectation 
value of the witness can be measured locally: Alice and Bob measure 
the expectation value of the projector and add their results with the
weights to receive the expectation value of the witness. Knowing the
value of $Tr(W\varrho)$ they can decide if the state is detected to be 
entangled, or not. The decomposition of the witness into projectors onto 
product vectors should be optimal, {\it i.e.} contain in a certain sense 
a possibly small number of terms.  

The paper is divided in four sections. In the first section we 
illustrate our proposal and our notation with a simple example: 
We consider a pure two-qubit state affected by white noise. 
The main parts of this paper are the extensions of this example:
In the second section we investigate the two-qubit scenario if the 
noise is not white. We also investigate the limits of our
scheme and the errors that may occur. 
The third section deals with extensions of the decomposition 
of the witness from the first section. We derive decompositions 
for witnesses in three-qubit case. With these operators GHZ-states 
and W-states can be detected. Eventually, we discuss how one can 
decompose witnesses for bipartite systems in higher dimensions.  
In the last section we apply our method to bound entangled states. 
We mainly discuss how some types of bound entanglement in $3\times 3$
and $2 \times 4$ systems can be detected.

\section{Two qubits with white noise}

We consider a setup that is intended to produce a particular pure state
$\KetBra{\psi},$ but due to the imperfections some noise is added. So, it 
produces a mixed state $\varrho$ of the form
\begin{equation}
\varrho(p,d) := p \KetBra{\psi} + (1-p)\sigma,
\label{rhodefinition}
\end{equation}
where we know that the noise added to the state $\Ket{\psi}$
is close to the totally mixed state, i.e.
\begin{equation}
\Vert \sigma - \frac{1}{4}\Eins \Vert \leq d.
\label{sigmanormbedingung}
\end{equation}
If $d=0$ the noise would be white, but in general this does not have to 
be the case. In the beginning, we do not make any assumption on  
$\Ket{\psi},$ but when we discuss the case $d>0$ we restrict ourselves 
to the important case that $\Ket{\psi}$ is a Bell state: 
$\Ket{\psi^+}=1/\sqrt{2} (\Ket{01}+\Ket{10}).$ Further we do not
make any restriction on $p$, {\it i.e.} we assume that it is unknown to us.
As a byproduct the witness will later donate us a possibility to 
determine $p.$ Our aim is to give a scheme to check whether $\varrho(p,d)$ 
is entangled or not.

Now we have to clarify our notation. We will denote the set of all density
matrices by $M$ and the set of separable matrices by $S.$ We can endow the 
space of all matrices 
with the scalar product $\HSNorm{A}{B}:=Tr(A(B^{\dagger}))$ and the 
corresponding 
Hilbert-Schmidt norm $\Vert A \Vert := \sqrt{Tr(A(A^{\dagger}))}.$  
For any $\varrho \in M$  and $r\geq 0$ we can define by
\begin{equation}
B(\varrho,r):=\{\varrho' \in M , \Vert \varrho'-\varrho \Vert \leq r \}
\end{equation}
the ball around $\varrho$ with radius $r.$ We denote by $B_{p,d}$ 
the ball which must include  $\varrho(p,d)$ for a given $p$ and $d.$ 
It is given by  $B_{p,d}=B(\frac{1-p}{4}\Eins+p\KetBra{\psi^+},(1-p)d).$ 
Finally we define $L$ as the line between $\Eins/4$ and $\KetBra{\psi};$
we have: $L=\{\varrho(p,0), \;\; p \in [0,1] \}.$

\subsection{Construction of the witness}

The optimal witness for an entangled $\varrho(p,0)$
is easy to construct \cite{witnesses2}. First, one has to compute the 
eigenvector corresponding to the negative
eigenvalue of $\varrho(p,0)^{T_B},$ then the witness is given by the 
partially transposed projector onto this eigenvector. 

If the Schmidt decomposition of $\Ket{\psi}$ is 
$\Ket{\psi} = a \Ket{01} + b \Ket{10}$  with $ a,b \geq 0,$ 
the spectrum of $\varrho(p,0)^{T_B}$ is given by 
$\{ (1-p)/4 + pa^2; (1-p)/4 + pb^2; (1-p)/4+p ab; 
(1-p)/4-pab \}.$  Therefore $\varrho(p,0)$ is entangled 
iff $p > 1/(1+4ab).$ The eigenvector corresponding
to the minimal eigenvalue $\lambda_-$ is given by
\begin{equation}
\Ket{\phi_-} = \frac{1}{\sqrt{2}}(\Ket{00}-\Ket{11}),
\end{equation}
and thus the witness $W_{0}$ is given by 
\begin{equation}
W_0 = \KetBra{\phi_{-}}^{T_B} 
    = \frac{1}{2}\left( \begin{array}{cccc}
1 & 0 & 0 & 0\\
0 & 0 &-1 & 0\\
0 &-1 & 0 & 0\\
0 & 0 & 0 & 1
\end{array}\right). 
\end{equation}
Note that this witness does neither depend on $p,$ nor on the 
Schmidt coefficients $a,b.$ It detects $\varrho(p,0)$ iff it
is entangled, since we have 
$Tr(\KetBra{\phi_{-}}^{T_B} \varrho(p,0)) = 
Tr( \KetBra{\phi_-} \varrho(p,0)^{T_B}) = \lambda_-.$
For our special case of white noise  $W_0$ has also the 
advantage that from $Tr(W_0 \varrho(p,0))\geq 0$ it follows 
that $\varrho(p,0)$ is separable. This is not a general 
property of witnesses, and indeed if the noise is not white 
this is not true anymore.

\subsection{Decomposition of the witness}

For an experimental setup  it is necessary to decompose the 
witness into operators which can be measured locally. Thus we 
need a decomposition into projectors onto product vectors 
of the form
\begin{equation}
W=\sum_{i=1}^{k} c_{i}\KetBra{e_{i}} \otimes 
\KetBra{f_{i}}. \label{pvdecomposition}
\end{equation}
Such a decomposition can be measured locally: Alice and Bob 
measure the expectation value of the 
$\KetBra{e_{i}} \otimes \KetBra{f_{i}}$ and add their results 
with the weights $c_i.$ One can construct such a decomposition in many 
ways, but it is reasonable to do it in a way which corresponds
to expenses of Alice and Bob which are as small as 
possible. There are several  possibilities to define an optimal 
decomposition: 

One possibility is to look for the optimal number of product vectors (ONP),
{\it i.e.} one can try to minimize $k$ in (\ref{pvdecomposition}). 
This optimization strategy looks very 
natural and has already been considered in the literature.
For $2 \times 2$ systems it was proven in \cite{sanpera98} 
that in general the ONP, which we denote as $k_-,$ is five. 
Also a constructive way for computing this optimal 
decomposition was given.  

What is the ``cost'' Alice and Bob have to pay when measuring
$W$ via such a decomposition? It is the number of measurements
they have to perform. When we talk about measurements here, we 
consider only von Neumann measurements. We do not look at POVMs since
their implementation would require additional ancilla systems.
One measurement on Alice's side in this sense consists of a 
choice of one orthonormal basis for Alice's Hilbert space.
For an particle with spin $s$ one may interpret this as the 
choice of a direction for a Stern-Gerlach-like apparatus. Alice 
sets up her device  in the desired direction and is able to 
distinguish between $2s+1$ different states.

For one local measurement Bob also has to choose an orthonormal 
basis in his Hilbert space; all together this yields one orthonormal 
product basis for both. Thus, if we are in a $N \times N$ system a 
term of the form
\begin{equation}
\sum_{k,l=1}^{N}c_{kl}\KetBra{A_k}\otimes\KetBra{B_l}
\label{lvnmdefinition}
\end{equation}
with $\BraKet{A_s}{A_t}=\BraKet{B_s}{B_t}=\delta_{st}$ can be 
measured with one collective setting of measurement devices 
of Alice and Bob. Alice and Bob can discriminate between the 
states $\Ket{A_kB_l},$ measure the probabilities of these states 
and add their results with the weights $c_{kl}$ using one
collective setting and some classical communication.
We call such a collective setting of measurement devices a local
von Neumann measurement (LvNM).

It is therefore reasonable to find a decomposition of the form
\begin{equation}
W=\sum_{i=1}^{m}\sum_{k,l=1}^{N}c^{i}_{kl}\KetBra{A^{i}_{k}}
\otimes\KetBra{B^{i}_{l}}
\label{problemdefinition}
\end{equation}
with $\BraKet{A^i_s}{A^i_t}=\BraKet{B^i_s}{B^i_t}=\delta_{st}$ and
an optimal number of devices' settings (ONS), {\it i.e.} a minimal $m.$  
In this sense  $m$ is the minimal number of measurements 
Alice and Bob have to perform. The construction of a decomposition of the form 
(\ref{problemdefinition}), and the determination of the minimal $m$
is the problem we want to solve in this section.

Please note that a decomposition like (\ref{pvdecomposition}) with 
the minimal $k_-$ (ONP) requires in general $k_-$ LvNMs because the 
out-coming vectors on Alice's side $\KetBra{e_i}$ do not have to 
be orthogonal.

We also would like to emphasize that a decomposition of the form 
(\ref{problemdefinition}) is more general than a decomposition
into a sum of tensor products of operators:
\begin{equation}
W=\sum_{i=1}^{m}\gamma_{i} A_i \otimes B_i.
\label{tensorproduktzerlegung}
\end{equation}
The decomposition (\ref{tensorproduktzerlegung}) has the 
advantage that Alice and Bob do not have to distinguish 
between some states, they only have to measure locally 
some expectation values of Hermitean operators. 
A decomposition like 
(\ref{problemdefinition}) can be written in the form 
of (\ref{tensorproduktzerlegung}) if for all $i$ the 
matrices $(c^i_{kl})$ are of rank one. In the 
following we will see that for qubit systems there is 
not a big difference between (\ref{problemdefinition})
and (\ref{tensorproduktzerlegung}). From the optimal 
decomposition in the sense of (\ref{problemdefinition})
we can derive a decomposition of the form 
(\ref{tensorproduktzerlegung}) where some of the 
operators are the identity ($\Eins$), so they do 
not require new measurement settings. For 
$N \times N$-systems we will see that it is straightforward 
to derive the optimal decomposition in the sense of 
(\ref{tensorproduktzerlegung}).

Our witness $W_0$ has a special form: It is a partially 
transposed projector $\KetBra{\psi}^{T_B}.$ It is therefore 
natural to look at the Schmidt decomposition of $\Ket{\psi}= 
\alpha\Ket{00}+\beta\Ket{11}.$ For our $W_0$ we have the 
special case  $\alpha=1/\sqrt{2}=-\beta,$ but we want to deal 
with the most general $\Ket{\psi}.$ 

When we compute the ONP-decomposition with the minimal $k_-$ according 
to \cite{sanpera98} we arrive at
\begin{equation}
\KetBra{\psi}^{T_B}=\frac{(\alpha+\beta)^2}{3}\sum_{i=1}^3 
\KetBra{ A'_i  B'_i} - \alpha \beta (\KetBra{01}+\KetBra{10}),
\end{equation}
where we have used the definitions
\begin{eqnarray}
\Ket{A'_1}& = & e^{i\frac{\pi}{3}} \cos(\theta) \Ket{0} 
                    + e^{-i\frac{\pi}{3}} \sin(\theta)\Ket{1} 
= \Ket{B'_1}
\nonumber \\
\Ket{A'_2}& = & e^{-i\frac{\pi}{3}} \cos(\theta) \Ket{0} 
                +e^{i\frac{\pi}{3}} \sin(\theta)\Ket{1}
= \Ket{B'_2}
\nonumber \\
\Ket{A'_3}& = & \Ket{A'_1}+\Ket{A'_2}
= \Ket{B'_3}
\nonumber \\
\cos(\theta)&=&\sqrt{\alpha/(\alpha+\beta)} \nonumber \\
\sin(\theta)&=&\sqrt{\beta/(\alpha+\beta)}.
\end{eqnarray}
This decomposition into five product vectors requires four correlated
settings for Alice and Bob. But we can measure $W_0$ with less settings:
If we  define the spin directions by $\Ket{z^+}=\Ket{0},\Ket{z^-}=\Ket{1},
\Ket{x^\pm}=\frac{1}{\sqrt{2}}(\Ket{0}\pm \Ket{1},
\Ket{y^\pm}=\frac{1}{\sqrt{2}}(\Ket{0}\pm i \Ket{1}$
we have the decomposition
\begin{eqnarray}
{\KetBra{\psi}^{T_B}}
&=& \alpha^2\KetBra{z^+ z^+}+\beta^2\KetBra{z^- z^-}+
\alpha\beta\left(\KetBra{x^+ x^+}+\right. 
\nonumber\\
& & \left.+\KetBra{x^- x^-}-\KetBra{y^+ y^-}-\KetBra{y^- y^+}\right)
\nonumber\\
&=&\frac{1}{4}\left(\Eins \otimes \Eins +\sigma_z\otimes \sigma_z 
+(\alpha^2-\beta^2)(\sigma_z\otimes\Eins+\Eins\otimes\sigma_z) \right. 
\nonumber \\
& &\left. +2\alpha\beta(\sigma_x\otimes \sigma_x+\sigma_y\otimes 
\sigma_y)\right).
\end{eqnarray}
This decomposition into six product vectors requires only a measurement 
of three settings: Alice and Bob have only to set up their Stern-Gerlach 
devices in the $x$-, $y$- and $z$-direction to measure $\KetBra{\psi}^{T_B}.$

Now we want to prove that three LvNM are really necessary.  
Our proof is a special case of a theorem about $N \times N$ systems 
we will show later. But in the two-qubit case the proof is 
particularly simple, and therefore we present it here separately.
\\
\\
{\bf Proposition 1.} In a two qubit system a decomposition of 
$\KetBra{\psi}^{T_B}$ of the form (\ref{problemdefinition}) 
requires at least three measurements.
\\
\emph{Proof.} Consider a decomposition requiring two measurements:
\begin{equation}
{\KetBra{\psi}}^{T_B}=
\sum_{i,j=1}^2 c^1_{ij} \KetBra{A^1_{i}} \otimes \KetBra{B^1_{j}}+
\sum_{i,j=1}^2 c^2_{ij} \KetBra{A^2_{i}} \otimes \KetBra{B^2_{j}}. 
\label{2x2decomposition}
\end{equation} 
With the help of a Schmidt decomposition as above we can write 
${\KetBra{\psi}}^{T_B}=\sum_{i,j=0}^{3} \lambda_{ij} \; 
\sigma_i \otimes \sigma_j$ with  
\begin{equation}
(\lambda_{ij})=
\left( \begin{array}{cccc}
\frac{1}{4}       &0            &0 & \frac{\alpha^2-\beta^2}{4}\\
0                 &\frac{\alpha\beta}{2} &0 & 0                \\
0                 &0            &\frac{\alpha\beta}{2} & 0     \\
\frac{\alpha^2-\beta^2}{4} &0            &0     & \frac{1}{4} 
\end{array}\right). 
\label{lambdamatrix}
\end{equation}
Note that the 3x3 submatrix in the right bottom corner is of rank 3.
Now we write any projector on the rhs of (\ref{2x2decomposition}) 
as a vector in the Bloch sphere: 
$\KetBra{A^1_{1}}=\sum_{i=0}^3 s^A_i \sigma_i$ is represented by
the vector $\vec{s}_{A^1_1}=(1/2,s^A_1, s^A_2,s^A_3)$ and 
$\KetBra{A^1_2}$ by $\vec{s}_{A^1_2}=(1/2,-s^A_1,-s^A_2,-s^A_3);$
$\KetBra{B^1_{1}}$ can be written similarly. If we expand the 
first sum on the rhs of (\ref{2x2decomposition}) in the 
($\sigma_i \otimes \sigma_j$) basis, the 3x3 submatrix in the right 
bottom corner is given by 
$(c^1_{11}-c^1_{12}-c^1_{21}+c^1_{22})
(s^A_1,s^A_2,s^A_3)^T(s^B_1,s^B_2,s^B_3).$ This matrix is of 
rank one.
The corresponding submatrix from the second sum on the rhs of 
(\ref{2x2decomposition}) is also of rank one and we arrive at a 
contradiction: No matrix of rank 3 can be written as a sum of two 
matrices of rank one.$\hfill\Box$

The idea of a generalization of this proposition to $N\times N$~
systems is straightforward: One expands both sides of an equation 
of the type (\ref{2x2decomposition}) in some product basis of the 
space of all operators. Then one tries to reach lower bounds for 
the rank of some submatrix of the coefficient matrix 
(\ref{lambdamatrix}) and upper bounds for the rank of the matrix 
corresponding to one LvNM. This gives a lower bound for the number 
of LvNMs.

\section{The case of non-white noise}

Our investigation of this case proceeds in two steps. First, we 
argue why the witness $W_0$ should also be used in this instance.
We  also show that under some circumstances, i.e. if 
$Tr(W_0 \varrho(p,d))$ is large enough, we can make a sure decision
that $\varrho(p,d)$ is separable. For the case that we cannot make 
a sure decision we derive analytical bounds and show numerical 
estimates for the error. This is done in the second step. 
 
We would like to remind the reader that if $d>0,$ we only consider 
maximally entangled states, so everywhere in this section we set 
$a=b=1/\sqrt{2}.$ 

\subsection{Properties of $W_0$}

Let us note some typical distances in $M$:
\\
\\
{\bf Remark 1. } (a) The states on $\partial M,$ the boundary 
of $M,$ closest to $\Eins/4$ have a distance of 
$1 / \sqrt{12} \approx 0.29.$ The states with the largest 
distance to $\Eins/4$ are just the pure states, for which the distance 
is $\sqrt{3}/2 \approx 0.87.$\\
(b) The states on $\partial S $ closest to $\Eins/4$
have also distance of $1/\sqrt{12}.$ 
\\
\emph{Proof.} These distances can be calculated simply by maximizing 
or minimizing $\Vert \Eins/4 - \varrho \Vert $ under some conditions,
for instance for (b) under the condition that 
$\mbox{det}(\varrho^{T_B})=0.$                     $\hfill \Box$

From part (a) it follows that we have to  assume that 
$d \leq 1/\sqrt{12}.$ Please note also  that the point where $L$
crosses the border of separability, $\partial S,$ is given by 
$\varrho(1/3,0)$ and this point has just the distance $1/\sqrt{12}$
from $\Eins/4,$ this means it is as close  as possible.
\\
\\
{\bf Remark 2.} For all $p$ the point on $\partial S$ 
closest to $\varrho(p,0)$ is given by $\varrho(1/3,0).$ 
\\
\emph{Proof.}
For $p<1/3$ we have mentioned it already. For $p>1/3$ we have 
a look at the set
\begin{equation}
\mathcal{N}(W_0)
:=\left\{ \varrho \in M ; Tr(W_{0}\varrho)=0 \right\}.
\label{nullstellenmenge}
\end{equation}
We have $\varrho(1/3,0) \in \mathcal{N}(W_0),$ and one can directly 
compute that for an arbitrary $N \in \mathcal{N}(W_0)$ and $\varrho(p,0)$ 
the relation $ \HSNorm{ \varrho(1/3,0)-\varrho(p,0)}{\varrho(1/3,0)-N} \; = 0$ 
holds.  Therefore $\mathcal{N}(W_0)$ and $L$ are orthogonal. From 
this and the properties of a witness operator the claim for 
$p>1/3$ follows. $\hfill \Box$
\\
\\
{\bf Remark 3.} The witness $W_0$  is the best possible witness 
in the following sense: For all $p$ and $d$ there is no other 
witness that detects a subset of $B_{p,d}$ as entangled which 
has a bigger volume than the subset of $B_{p,d}$ which is 
detected as entangled by $W_0.$
\\
\emph{Proof.} From the second remark it follows that if 
$B_{p,d} \cap S = \emptyset$ the witness $W_{0}$ detects the whole 
ball, independent of $d.$

Thus,  we can assume that  $p$ has a value such that  $B_{p,d}$ contains 
separable and entangled states and that there exists a witness 
$W',$ which detects a bigger volume of $B_{p,d}$. Then we look at 
the set $\mathcal{N}(W'),$ defined analogous to 
(\ref{nullstellenmenge}). If $p < 1/3$ the minimal distance between 
matrices of $\mathcal{N}(W')$ and $\varrho(p,0)$ must be smaller 
than the minimal distance between $\mathcal{N}(W_0)$ and 
$\varrho(p,0),$  which is 
$\Vert \varrho(p,0)-\varrho(1/3,0)\Vert.$ This means that 
there exist entangled states (and therefore states on $\partial S$) 
which are closer to $\varrho(p,0)$ than $\varrho(1/3,0)$ and 
we have a contradiction to Remark 2. If $p \geq 1/3$ 
the minimal distance between matrices of $\mathcal{N}(W')$ 
and $\varrho(p,0)$ must be bigger than the distance between 
$\mathcal{N}(W_0)$ and $\varrho(p,0),$ but then $W'$ must 
``detect'' the separable state $\varrho(1/3,0),$ which is a 
contradiction. $\hfill \Box$

The critical reader may ask at this point why we are so innocent and 
use the term of a ``volume'' in $M.$ It seems that it is difficult to say 
something about volumes in our norm, since we are not in the 
$\Bbb{R}^n$ with the Euclidean norm \cite{volumes,volumes2}. But we are not 
too far away from the $\Bbb{R}^n:$ We can write any density matrix as 
$\varrho=\sum_{i=0}^{15}\mu_i G_i$ where $\mu_i \in \mathbb{R},$   
$G_0 \sim \Eins$ and the $G_1,...,G_{15}$ are the traceless generators of 
the $SU(4).$ The $G_i$ can be normalized in a way that they form
an orthonormal basis: $\HSNorm{G_i}{G_j}=\delta_{ij}.$ We will explain 
and use this decomposition in greater detail later. Here we only point 
out that $\Vert \varrho \Vert^2 = \sum_{i=0}^{15} \mu_i^2$ and so our 
norm in the space of Hermitean operators  is just the Euclidean norm 
in  $\Bbb{R}^{16}.$ $M$ corresponds to a subset of a $15$-dimensional
hyperplane since $\mu_0$ is fixed, and the usual formulas for volumes 
of balls can be applied.

Now having proven that it is reasonable to use $W_0,$ one may ask
what the expectation value of $W_0$  tells us for the case $d>0.$
From the definition of a witness operator it follows that if the 
expectation value is negative we can be sure that $\varrho(p,d)$ is 
entangled. On the first view it seems that if it is positive 
we can not make a sure decision if 
$\varrho(p,d)$ is entangled or not. But if  $Tr(W_0\varrho(p,d))\gg 0$ 
this means that $\varrho(p,d)$ is far away from $\mathcal{N}(W_0)$ 
and since we know that $\varrho(p,d) \in B_{p,d}$ for some $p,$ the state 
$\varrho(p,d)$ must be separable. So there must exist a $\tau(d)$ such 
that from $Tr(\varrho(p,d) W_0) \geq \tau$ it follows that $\varrho(p,d)$ 
is separable.
We can directly compute $\tau,$ but first we introduce some new
definitions.

If we measure  $Tr(\varrho W_0)=\alpha$ the expectation value 
$\alpha$ tells us that $\varrho$ is in some hyperplane cutting $M.$ 
This hyperplane is orthogonal to $L,$ as we have shown in the 
proof of remark 2.
It intersects this line at some $\varrho(q,0)$ and so we denote this 
plane by $\mathcal{P}(q).$ For instance an essential part of the 
proof of Remark 2 was the statement that 
$\mathcal{P}(1/3)=\mathcal{N}(W_0).$ The connection between the 
expectation values of $W_0$ and the planes $\mathcal P$
is given by:
\begin{equation}
Tr(\varrho W_0)=\alpha \;\; \Leftrightarrow \;\; 
\varrho \in \mathcal P(q) 
\;\; \mbox{with} \;\;  
q = \frac{1}{3}-\frac{4}{3}\alpha. 
\label{epspumrechnung}
\end{equation}
Since we do not deal with general $\varrho,$ but with some $\varrho(p,d),$
it will be useful to consider the set of all possible $\varrho(p,d).$ This is
\begin{equation}
K(d):=\bigcup_{p\in[0,1]} B_{p,d}.
\label{kdefinition}
\end{equation}
We can now define the intersections 
\begin{eqnarray}
SP(q)    &:=& S\cap\mathcal{P}(q) ,\nonumber \\
KP(q,d)  &:=& K(d)\cap\mathcal{P}(q), \nonumber \\
BP(q,p,d)&:=& B_{p,d}\cap\mathcal{P}(q), \nonumber \\
XP(q)   &:=& 
B(\frac{1}{4}\Eins,\frac{1}{\sqrt{12}})\cap\mathcal{P}(q).
\end{eqnarray}
$SP$ is the set of all separable states which belong to one 
possible expectation value. $KP$ is the set of all 
possible $\varrho(p,d)$ yielding the same expectation value.
It is clear that $BP(q,p,d) \subseteq KP(q,d).$ Since we 
know already from Remark 1 that the states in 
$ B(\Eins/4,1/\sqrt{12})$ are separable, we can conclude 
that $XP(q) \subseteq SP(q).$ 
For the sake of notational simplicity we often suppress the 
parameters $q,p,d.$

The strategy of computing $\tau$ is now clear: We have to find the 
values of $q$ for which $KP(q)\subseteq SP(q)$ holds. For the 
corresponding expectation values of $W_0$ our knowledge that 
$\varrho(p,d)\in K$ guarantees us that $\varrho(p,d)$ is separable.
Since it is difficult to characterize $SP$ we replace it by $XP.$ 
This replacement is justified later. 
\\
\\
{\bf Proposition 2.} Let
\begin{equation}
\tau=\frac{1}{4}-d^2-
         \sqrt{\left(\frac{1}{12}-d^2 \right)\left(\frac{3}{4}-d^2\right)}.
\label{epsmaxdefinition}
\end{equation}
Then we have:
\begin{equation}
Tr(\varrho(p,d) W_{0})\geq \tau 
\;\; \Rightarrow \;\;  
\varrho(p,d) \mbox{ is separable.}
\label{epsmaxeigenschaft}
\end{equation} 
Furthermore: The $\tau$ defined in (\ref{epsmaxdefinition}) is the
minimal $\tau$ with the property (\ref{epsmaxeigenschaft}), i.e. 
for all $ 0\leq \tau' < \tau $ there exists an entangled state 
$\varrho (p,d)$ with $Tr(\varrho(p,d) W_{0}) = \tau'.$
\\
\emph{Proof.} 
The idea of the proof is as described above: First, we compute the parameter 
$q_-,$ such that for $q\leq q_-$ we have $KP(q)\subset SP(q).$ 
Via (\ref{epspumrechnung}) we arrive at $\tau.$ Finally we show 
that for $q > q_-$ we have $KP(q)\not\subset SP(q).$

$KP$ is just like $BP$ a ball in the hyperplane $\mathcal{P}$. 
The radius of $BP$ is determined by 
$r^2(BP)=(1-p)^2 d^2-\Vert \varrho(q, 0)-\varrho(p,0) \Vert^2 =
(1-p)^2 d^2-(3/4)(q-p)^2.$ 
Maximizing this over all $p$ yields
\begin{equation}
r^2(KP) = \frac{3d^2}{3-4d^2}(1-q)^2.
\end{equation}
The  ball $XP$ has the squared radius
\begin{equation}
r^2(XP) = \frac{1}{12}-\frac{3}{4} (q)^2.
\end{equation}
If we choose $q$ small enough, it follows from 
$XP(q) \subset SP(q)$ that $KP(q)\subset XP(q) \subset SP(q).$ 
So we can determine $q_-$ by the equation  
$r^2(KP(q_-)) = r^2(XP(q_-)),$  apply (\ref{epspumrechnung})
and arrive at $\tau.$

Why can we not use a smaller $\tau$? The reason is that for any 
$q \in [0,1/3]$ the ball $XP(q)$ contains at least one state of 
$\partial S.$ This means that from $KP(q)\not\subseteq XP(q)$ it 
follows that $KP(q)\not\subseteq SP(q).$
To see the existence of a state $\in \partial S$ in $XP(q)$
first note that if we take our first Bell state 
$\Ket{\psi^+}=(1/\sqrt{2})(\Ket{01}+\Ket{10})=:
(1/\sqrt 2)(\Ket{0' 0'}+\Ket{1' 1'})$
and another Bell state
$\Ket{\phi^+}=\frac{1}{\sqrt 2}(\Ket{00}+\Ket{11})$
we have: 
\begin{equation}
\HSNorm{(\frac{1}{4}\Eins- \KetBra{\psi^+})}{(\frac{1}{4}\Eins- 
\KetBra{\phi^+})}
 = -\frac{1}{4}<0.
\end{equation} 
We can transform the bases 
$\Ket{0'0'} \rightarrow \Ket{00}, \Ket{1'1'} \rightarrow \Ket{11},$ 
in a continuous way and thereby construct a continuous map
\begin{equation}
\gamma: [0,2] \rightarrow M; \;\; t \mapsto \gamma(t)
\end{equation}
with $\gamma(0)=\KetBra{\psi^+}$ and $\gamma(2)=\KetBra{\phi^+}$ and 
for all $t,$ $\gamma(t)$ is a maximally entangled state. There must exist 
a $t_0$ (and without loosing generality we can set $t_0=1$) with
\begin{equation}
\HSNorm{(\frac{1}{4}\Eins-\KetBra{\psi^+})}{(\frac{1}{4}\Eins-\gamma(1))} = 0.
\end{equation}
This equation implies that $\gamma(1)\in \mathcal{P}(q=0).$
The map $\gamma$ induces another continuous map by
\begin{equation}
\eta: [0;1] \rightarrow S; \;\; 
t \mapsto \eta(t)=\frac{2}{3}\frac{1}{4}\Eins +\frac{1}{3}\gamma(t).
\end{equation}
Since $\gamma(t)$ is a maximally entangled state $\eta(t)$ is a
point on $\partial S.$ Furthermore all $\eta(t)$ have a distance 
of $1/\sqrt{12}$ to the state  $\Eins/4,$ i.e. 
$\eta(t) \in B(\Eins/4,1/\sqrt{12}).$
Since $\eta(0) \in \mathcal P(1/3)$ and $\eta(1) \in \mathcal P(0),$ 
for any $q \in [0,1/3],$ $\mathcal P(q)$ contains one $\eta(t_q)$  
and therefore $XP(q)$ contains a state on $\partial S.$ $\hfill \Box$

If one knows the value of $p$ one can sharpen the statement of the
last proposition:
\\
\\
{\bf Proposition 3.} Let $p$ be fixed, $\varrho(p,d) \in B_{p,d}$ and
\begin{equation}
\vartheta := \frac{1}{4}-\frac{1}{24p}-\frac{3p}{8}+\frac{(1-p)^2 d^2}{2p}.
\label{deltamaxdefinition}
\end{equation}
Then we have:
\begin{equation}
Tr(\varrho(p,d) W_{0})\geq \vartheta 
\;\; \Rightarrow \;\;  
\varrho(p,d) \mbox{ is separable.}
\label{deltamaxeigenschaft}
\end{equation} 
Furthermore: The $\vartheta$ defined in (\ref{deltamaxdefinition}) is the
minimal $\vartheta$ with the property (\ref{deltamaxeigenschaft}) i.e. for all
$ 0\leq \vartheta' < \vartheta $ there exists an entangled state 
$\varrho (p,d)$ with $Tr(\varrho(p,d) W_{0}) = \vartheta'.$
\\
\emph{Proof.} The proof is essentially the same as the proof of Proposition 2.
One just has to replace $K$ by $B_{p,d}$ and therefore $KP(q)$ by $BP(q).$
$\hfill \Box$

\subsection{Error estimates}

Consider the case that one has measured $Tr(W_0 \varrho(p,d))=\alpha$
with $\alpha \in [0,\tau].$ For such expectation value one cannot
make a sure decision whether $\varrho$ is entangled or not. Nevertheless
one can make a decision, if one accepts to make some error. Here we 
want to estimate the probability of making an error.  First we give 
some analytical bounds on the error, then we perform some numerical
simulations to rate the error.

For an estimate of an error, we have to make one further 
assumption on $\varrho(p,d):$ We have to assume some probability 
distribution. We always assume that for all $p$ and $d,$ 
$\varrho(p,d)$ is uniformly distributed in $B_{p,d}.$

\subsubsection{Analytical estimates}
Our analytical estimation scheme relies on the fact that all states
in the ball $B(\Eins/4,1/\sqrt{12})$ are separable. Thus we know at 
least some separable states yielding expectation values 
$\alpha \in [0,\tau].$ If we assume that $\varrho(p,d)$ 
is separable, this gives us an upper bound on the probability of 
the error. 
 
For an illustration of this idea let us first assume that we know 
the fixed value of $p$ and we have measured 
$Tr(W_0 \varrho(p,d))=\alpha.$ Via (\ref{epspumrechnung}) we can 
compute the $q$ such that $\varrho(p,d) \in \mathcal{P}(q).$
We know then that $\varrho(p,d) \in BP(q).$ If we now assume 
that $\varrho(p,d)$ is separable, the probability of guessing 
right is given by the ratio of the volumes
\begin{equation}
e_{r} = \frac{\mbox{vol}(SP(q))}{\mbox{vol}(BP(q))},
\end{equation}
and the probability of being wrong is bounded by
\begin{equation}
e_{w} \leq E_{w}=
\frac{\mbox{vol}(BP(q))-\mbox{vol}(XP(q))}
{\mbox{vol}(BP(q)}. 
\end{equation}
These terms can be calculated with the standard formulas for balls 
in higher dimensions: 
$\mbox{vol}(BP(q))=(\pi^7/7!)((1-p)^2d^2-3/4(p-q)^2)^7$ and 
$\mbox{vol}(XP(q)) =(\pi^7/7!)(1/12-3/4q^2)^7.$

Since we do not know the value of $p$ we have to maximize $E_w$
over all $p$, then we arrive at
\begin{equation}
E_-= \sup_{p\in[0,1]} E_w = 1 - 
\frac{\left( \alpha (\alpha - \frac{1}{2}) (d^2-\frac{3}{4})\right)^7}
{\left( d (\alpha + \frac{1}{2}) \right)^{14}}. 
\label{omegadefinition}
\end{equation}
This function is plotted together with numerical results
in Fig. 1. 

Let us mention that
with the same method as above one can also estimate the error for other 
scenarios, for instance if one generally assumes that for 
$Tr(W_0 \varrho(p,d))\geq 0$ $\varrho$ is separable without looking
at the precise expectation value; we will not discuss further this 
direction here, however.

\subsubsection{Numerical calculations}
One can investigate the properties of the witness also with numerical 
calculations. For this purpose one can generate random matrices 
\cite{volumes2} 
and compare the result of the witness with the PPT criterion, which is 
a necessary and sufficient criterion of separability for $2 \times 2$ 
systems \cite{perescrit}. We have generated a sample of 50000 random matrices  
(in Hilbert-Schmidt norm) in the ball $B(\Eins/4,1/\sqrt{12}).$
By scaling we have retrieved then a set of 50000 random matrices 
$\varrho(p,d)$ for all $p$ and $d.$ With these matrices we have obtained 
our numerical estimates.

First, we investigated the error estimate with $E_-$ 
from Eq. (\ref{omegadefinition}). For this purpose we computed $e_w$
and then $e_-$ as the supremum of $e_w$ over all $p.$ This 
is shown in Fig. 1.

\begin{figure}[h]
\centerline{\psfig{figure=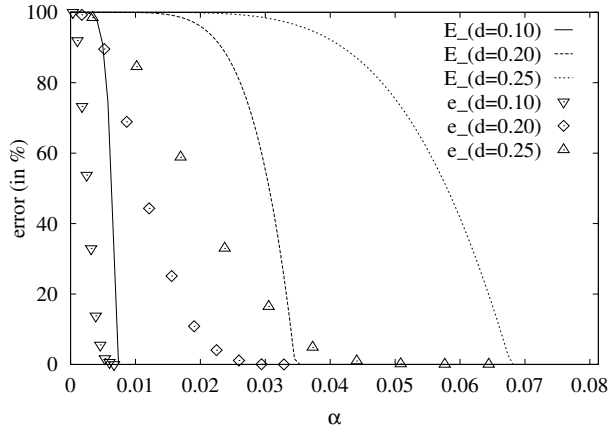,width=0.7\textwidth}}
\caption{ Comparison between the numerical estimate ($e_-$) and 
the analytical bound ($E_-$) as functions of the expectation value.
The three curves are for three different values of $d.$}
\end{figure}

One may ask how big the error is, if one generally concludes 
from $Tr(W_0\varrho(p,d)) \geq 0$ that $ \varrho(p,d) \in S.$
We have estimated the error (maximized over all $p$). The 
result is given in Fig. 2.

\begin{figure}[h]
\centerline{\psfig{figure=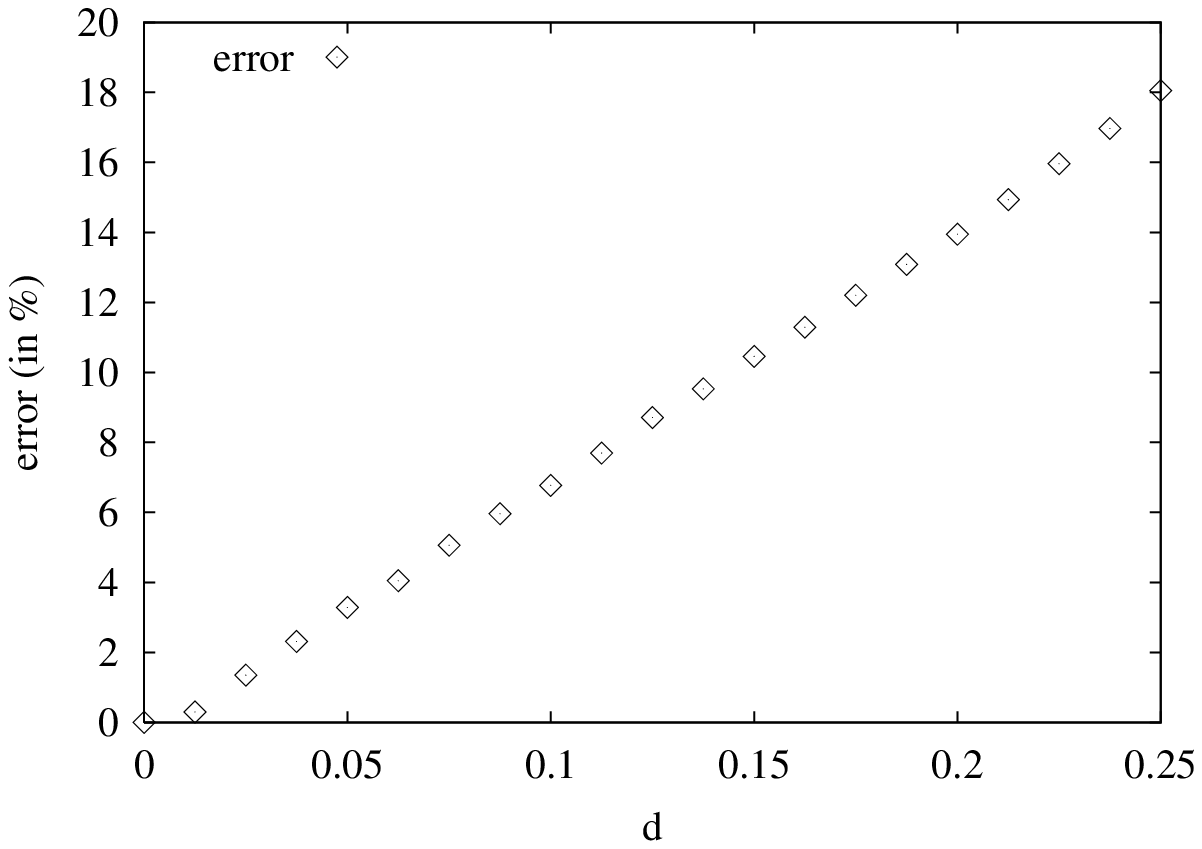,width=0.7\textwidth}}
\caption{Probability of making an error (maximized over all $p$) when
assuming $Tr(W_0\varrho(p,d))>0 \Rightarrow \varrho(p,d) \in S$ as a 
function of $d.$} 
\end{figure}

\section{Applications for 3-qubit and $N \times N$ systems}

First, we consider three-qubit systems. For these systems two types 
of tripartite entanglement are known: The GHZ-states and the 
W-states \cite{duer2000}.
Also families of witnesses for these states are known \cite{acin2001} 
and we show how to decompose them locally. 
Then we say something  about a decomposition of a projector 
in $N \times N$ systems.
 
\subsection{Three qubits}

Three qubits can be entangled in different ways: They might be 
separable, biseparable of fully tripartite entangled \cite{acin2001}.
The genuine threepartite entanglement consists of two classes:
The GHZ-class and the W-class \cite{duer2000}. 

Witnesses for detection of GHZ-type states and W-type states have
also been constructed in \cite{acin2001}. Here we want to show that
these witnesses can be decomposed within our scheme. 

For the GHZ-class a witness is given by
\begin{equation}
W_{GHZ}=\frac{3}{4}\Eins-\KetBra{GHZ},
\label{ghzwitnessdefinition}
\end{equation}
where $\Ket{GHZ}$ is a pure state of the GHZ-class: 
$\Ket{GHZ}=1/\sqrt{2}(\Ket{000}+\Ket{111}).$ If $\varrho$ is a 
mixed state with $Tr(\varrho W_{GHZ})<0$ the state $\varrho$ belongs 
to the GHZ-class. A decomposition of $W_{GHZ}$ can be achieved with 
similar calculations as above. The result is:
\begin{eqnarray}
W_{GHZ}&=&
\frac{1}{8}\left(
5 \cdot\Eins\otimes\Eins\otimes\Eins-
\Eins\otimes\sigma_z\otimes\sigma_z-
\sigma_z \otimes \Eins \otimes \sigma_z- 
\sigma_z \otimes \sigma_z \otimes \Eins - 
\right. 
\nonumber \\
& & 
\left.-
\sigma_x \otimes \sigma_x \otimes \sigma_x +
\sigma_x \otimes \sigma_y \otimes \sigma_y +
\sigma_y \otimes \sigma_y \otimes \sigma_x +
\sigma_y \otimes \sigma_x \otimes \sigma_y \right).
\nonumber \\
&&
\label{ghzwitnesszerlegung}
\end{eqnarray}
This witness can be measured with five collective measurement settings: 
Alice, Bob and Charly have to perform correlated measurements in the
$z$-$z$-$z$, $x$-$x$-$x$, $x$-$y$-$y$,$y$-$x$-$y$ and the $y$-$y$-$x$ direction.

For the W-states several witnesses are known. One example is the 
operator 
\begin{equation}
W_{W1}=\frac{2}{3}\Eins-\KetBra{W},
\end{equation}
where $\Ket{W}$ is now a pure state of the W-class: 
$\Ket{W}=1/\sqrt{3}(\Ket{100}+\Ket{010}+\Ket{001}.$ This witness 
detects states belonging to the W-class and the GHZ-class, i.e. it's 
expectation value is positive on all biseparable and fully separable 
states. A decomposition is given by
\begin{eqnarray}
W_{W1}&=&\frac{1}{24}
\bigl(
13 \cdot \Eins \otimes \Eins \otimes \Eins-
\sigma_z \otimes \Eins \otimes \Eins -
\Eins \otimes \sigma_z \otimes \Eins -
\Eins \otimes \Eins \otimes \sigma_z +
\nonumber \\
&&
+\sigma_z \otimes \sigma_z \otimes \Eins + 
\sigma_z \otimes \Eins \otimes \sigma_z+ 
\Eins \otimes \sigma_z \otimes \sigma_z +
3 \cdot \sigma_z \otimes \sigma_z \otimes \sigma_z 
\bigr)
\nonumber \\
&& -\frac{1}{12} \left(
\Eins \otimes \sigma_x \otimes \sigma_x +
\Eins \otimes \sigma_y \otimes \sigma_y +
\sigma_z \otimes \sigma_x \otimes \sigma_x +
\sigma_z \otimes \sigma_y \otimes \sigma_y +
\right.
\nonumber \\
&&
\sigma_x \otimes \Eins \otimes \sigma_x+ 
\sigma_y \otimes \Eins \otimes \sigma_y+ 
\sigma_x \otimes \sigma_z \otimes \sigma_x+ 
\sigma_y \otimes \sigma_z \otimes \sigma_y+ 
\nonumber \\
&& \left.
\sigma_x \otimes \sigma_x \otimes \Eins + 
\sigma_y \otimes \sigma_y \otimes \Eins + 
\sigma_x \otimes \sigma_x \otimes \sigma_z + 
\sigma_y \otimes \sigma_y \otimes \sigma_z 
\right).
\end{eqnarray}
Although this decomposition is a little bit longer, only seven
correlated measurements are necessary.

Another witness for W-class states is given by
\begin{equation}
W_{W2}=\frac{1}{2}\Eins-\KetBra{GHZ}.
\label{w2witness}
\end{equation}
This witness can be measured locally with the same decomposition as
(\ref{ghzwitnesszerlegung}). 
It also can serve for a detection of states of the type 
$(1-p)\Eins/8+p\KetBra{W},$ as explained in \cite{acin2001}.

\subsection{$N\times N$ systems}
Now we want to generalize our results to higher dimensions. First we 
consider $N \times N$ systems, and the end of this section we make 
some remarks about  $N \times M$ systems. 

A witness for an entangled state $\varrho $ with a non positive partial 
transpose can be constructed just like in the two-qubit case. First,
one computes one eigenvector corresponding to one negative eigenvalue
of $\varrho^{T_B}.$  The partially transposed projector onto this 
vector is an entanglement witness. We want to decompose such witnesses 
for NPT states in this section. We only look at projectors, the partial 
transposition can be performed later.

Our discussion proceeds as follows: After explaining our notation 
we construct a decomposition of a projector onto a state with Schmidt 
rank $l$ using about $2l$ measurements. This decomposition is a 
generalization of the decomposition for the two qubit case. It is not 
clear whether this decomposition is optimal. Then we derive a lower 
bound for the number of measurements needed if the Schmidt rank $l$ is 
maximal. We show that if $l=N$ at least $l+1$ measurements are necessary.

We first explain some notational and technical details (the reader
should consult \cite{schlienz95} for more explanations). We denote the 
real vector space of all Hermitean operators on $\HH_A$ by $\HS_A.$ 
In this space one can use the orthogonal basis 
$\{\Eins, G^A_i, i=1...N^2-1\}$ where the $G^A_i$ are the traceless 
generators of the $SU(N),$ normalized to $Tr(G_i^2)=1.$ For $N=2$ they 
are the Pauli matrices, for $N=3$ the Gell-Mann matrices, etc. We can define 
$G^A_0:=\Eins$ and can expand every projector (and any other element 
of $\HS_A$) in this basis:
\begin{equation}
\KetBra{\phi}=\sum_{i=0}^{N^2-1} f_i G^A_i,
\label{blochdarstellung}
\end{equation}
where the entries of the Bloch vector $f_i$ are real, $f_0 = 1/N,$ 
and from the fact that $\KetBra{\phi}$ is a pure state 
it follows that
\begin{equation}
\sum_{i=1}^{N^2-1}f_i^2=1-\frac{1}{N}. 
\label{purestatecondition}
\end{equation}
We sometimes write $(f_0,...,f_{N^2-1})=(f_0,\vec{f})=\hat{f}.$ It 
is easy to see that an operator described by $\hat{g}$ is a projector
onto a vector orthogonal to $\Ket{\phi}$ if and only if $\vec{g}$ fulfills
(\ref{purestatecondition}) and 
\begin{equation}
<\vec{f},\vec{g}>:=\sum_{i=1}^{N^2-1}f_i g_i = -\frac{1}{N}.
\label{orthogonalitaetsbedingung}
\end{equation}
One can also expand any projector (as every operator) on 
$\HH_A\otimes\HH_B$ as
\begin{equation}
\KetBra{\psi}=\sum_{i,j=0}^{N^2-1}\lambda_{ij} G^A_i\otimes G^B_j
\label{lambdaijgleichung}
\end{equation}
since the $G^A_i\otimes G^B_j$ form a product basis of the space
$\HS=\HS_A\otimes\HS_B.$

Before we show  our decomposition please note that is easy to 
decompose any operator $A \in \HS$ into $N^2$ local measurements.
One can always write
\begin{equation}
A=\sum_{i,j=0}^{N^2-1}\mu_{ij} G^A_i\otimes G^B_j =
\sum_{i=0}^{N^2-1} G^A_i \otimes 
\left(\sum_{j=0}^{N^2-1} \mu_{ij} G^B_j\right)
\label{einfachezerlegung} 
\end{equation}  
to obtain such a decomposition. This is also a decomposition of 
the form (\ref{tensorproduktzerlegung}).
\\
\\
{\bf Theorem 1.} Let $\KetBra{\psi}$ be a projector onto a state with
Schmidt rank $l.$  If $l$ is even, $\KetBra{\psi}$ can be decomposed  into
$2l-1$  local measurements. If $l$ is odd, $\KetBra{\psi}$ can be 
decomposed  into $2l$ local measurements.\\
\emph{Proof.} If we have $\Ket{\psi}= \sum_{i=1}^l s_i\Ket{ii}$ we
can write
\begin{equation}
\KetBra{\psi}=\sum_{i=1}^{l}s_i^2\KetBra{ii}+
\sum_{i,j=1,i < j}^{l}s_i s_j K(i,j)
\label{projektorzerlegung}
\end{equation} 
with $K(i,j)=\Ket{ii}\Bra{jj}+\Ket{jj}\Bra{ii}.$ The first sum 
corresponds to one measurement and every of the $l(l-1)/2$ terms 
of the second sum can be decomposed by defining for every
$K(i,j)$ the directions 
$\Ket{X^\pm_{i,j}}=\frac{1}{\sqrt{2}}(\Ket{i}\pm \Ket{j},
\Ket{Y^\pm_{i,j}}=\frac{1}{\sqrt{2}}(\Ket{i}\pm i \Ket{j}$
and writing:
\begin{eqnarray}
K(i,j)&=&\KetBra{X^+_{i,j} X^+_{i,j}}+\KetBra{X^-_{i,j} X^-_{i,j}}-\nonumber\\ 
      & & -\KetBra{Y^+_{i,j} Y^+_{i,j}}-\KetBra{Y^-_{i,j} Y^-_{i,j}}), 
\label{kreuztermzerlegung} 
\end{eqnarray}
as we have done before for $2\times2$ systems. This corresponds to $2$ 
measurements for each $K(i,j).$

The idea is now to sum up the terms from (\ref{kreuztermzerlegung}) 
for different $K(i,j)$ and $K(m,n)$ in a way that the terms from 
different $K(i,j)$ and $K(m,n)$ can be measured with one measurement.

Let us first consider the case that $l$ is even. We have $l(l-1)/2$
index pairs $(i,j).$ These pairs can be grouped into $l-1$ sets 
of $l/2$ pairs in a way that in every set every index $1 \leq i \leq N$ 
appears exactly in one pair. For instance for $l=4$ the $3$ sets may 
be defined as $\{(1,2),(3,4)\},\{(1,3),(2,4)\},\{(1,4),(2,3)\}.$ 
If we look at the $l/2$ $K(i,j)$ belonging to one set, 
the corresponding 
vectors $\Ket{X^\pm_{i,j}}$  are mutually orthogonal, they form an 
orthogonal 
basis of $\HH_A$ and $\HH_B.$ So all these vectors can be viewed as 
eigenvectors of some Hermitean operator on $\HH_A$  and $\HH_B$ and 
can be measured with one measurement. The vectors  $\Ket{Y^\pm_{i,j}}$  
can also be measured with one measurement. So we need $2$ measurements 
for one set and  $2(l-1)$ measurements for all $K(i,j).$ Finally we 
need one measurement for the first sum on the rhs of 
(\ref{projektorzerlegung}) and this completes the proof for even $l.$

If $l$ is odd, we can similarly group the $l(l-1)/2$ index 
pairs into  $l$ sets of $(l-1)/2$ pairs. This time in every set 
every index appears at most one time, one index is missing in every set 
and every index is missing in exactly one set. As before 
we need $2$ measurements for one set and therefore $2l$ measurements 
for all $K(i,j).$ For the first sum on the rhs of (\ref{projektorzerlegung}) 
we do not need another measurement since  we can put vector $\Ket{i}$  to the 
set of index pairs where $i$ is missing. $\hfill\Box$

Now we want to give a lower bound for the number of required 
measurements for a projector. This bound is based on the same idea
as the proof of Proposition 2 and needs three lemmata. In Lemma 1 
we give a lower bound for the rank of some matrix of the form 
(\ref{lambdamatrix}) in $N \times N$-systems. In the Lemmata 2 and 3
we show that the matrix coming from one measurement has a low rank.
Together this proves our bound. 
\\
\\
{\bf Lemma 1.} If $\Ket{\psi} \in \HH_A\otimes\HH_B $ has the full 
Schmidt rank $N$ then the matrix $(\lambda_{ij})$ in 
(\ref{lambdaijgleichung}) has the full rank $N^2$.
\\
\emph{Proof.} First, notice that the rank of $(\lambda_{ij})$ is 
independent of the choice of the basis $G^A_i\otimes G^B_j.$ If 
one has another basis $H^A_i\otimes H^B_j$ with 
$G^A_i=\sum_l a_{il}H^A_l$ and $G^B_j=\sum_r b_{jr}H^B_r$ the new matrix
of coefficients is given by 
$(\lambda'_{lr})=\sum_{i,j} a^T_{li}\lambda_{ij} b_{jr}$ 
and since the matrices $(a_{il})$ and $(b_{jr})$ have full rank 
the matrix $(\lambda'_{ij})$ has the same rank as $(\lambda_{ij}).$

Now we simply  construct an orthonormal product basis of $\HS$ where 
$(\lambda'_{ij})$ is diagonal and the diagonal elements do not vanish. 
Starting from the Schmidt-decomposition  
$\Ket{\psi}= \sum_{i=1}^N s_i\Ket{ii}$ we define on $\HH_A,$ as 
well as on $\HH_B:$
\begin{eqnarray}
P_k&=&\KetBra{k},\qquad  \qquad \qquad \qquad 1\leq k \leq N \\
Q_{jk}&=&\frac{1}{\sqrt{2}}(\Ket{j}\Bra{k}+\Ket{k}\Bra{j}), 
\quad 1\leq j < k \leq N \\
R_{jk}&=&\frac{i}{\sqrt{2}}(\Ket{j}\Bra{k}-\Ket{k}\Bra{j}),
\quad 1\leq j < k \leq N. 
\end{eqnarray}
These $N^2$ operators form an orthonormal basis of $\HS_A$ (resp. $\HS_B$), 
denoted by $H_i^A$ (resp. $H_i^B$), and if one computes
\begin{equation}
\lambda'_{rs}=
\sum_{\alpha,\beta=1}^{N}s_{\alpha}s_{\beta} 
\Bra{\alpha}H_r^A\Ket{\beta}\Bra{\alpha}H_s^B\Ket{\beta}
\end{equation}
one can directly verify that $(\lambda'_{rs})$ is in the basis 
$H_r^A \otimes H_s^B$ diagonal and has the full rank. $\hfill\Box$
\\
\\
{\bf Corollary 1.}
If $\Ket{\psi} \in \HH_A\otimes\HH_B $ has the 
Schmidt rank $l$ then the matrix $(\lambda_{ij})$ in 
(\ref{lambdaijgleichung}) has the rank $l^2$.
\\
\emph{Proof.} The proof is essentially the same as the proof 
of Lemma 1. We can view $\Ket{\psi}$ as a vector in a 
$l \times l$-system. $\hfill\Box$
\\
\\
{\bf Corollary 2.}
If $\Ket{\psi} \in \HH_A\otimes\HH_B $ has the Schmidt 
rank $l$ then a decomposition in the sense of 
(\ref{tensorproduktzerlegung}) requires $l^2$ Hermitean 
operators for every party.
\\
\emph{Proof.} If one would need less, this would be a direct
contradiction to Lemma 1 and Corollary 1. Please note that we have 
already computed this decomposition -- see 
(\ref{einfachezerlegung}). $\hfill\Box$
\\
\\
{\bf Lemma 2.}  Let $\vec{v}_1,...,\vec{v}_r \in \Bbb{R}^n$ be some 
vectors obeying the equations
\begin{equation}
<\vec{v}_i,\vec{v}_j>=C\neq0 \qquad \forall \; i \neq j. 
\label{kegelgleichung}
\end{equation}
$\vec{v}_r$ should be uniquely defined by $\vec{v}_1,...,\vec{v}_{r-1}$ 
and the equations (\ref{kegelgleichung}) while $\vec{v}_{r-1}$ should 
not be uniquely defined by $\vec{v}_1,...,\vec{v}_{r-2}$ and the equations 
(\ref{kegelgleichung}). Then we have 
\begin{equation}
\vec{v}_r \in \mbox{Lin}(\vec{v}_1,...,\vec{v}_{r-1})
\end{equation}
and 
\begin{equation}
\mbox{dim}(\mbox{Lin}(\vec{v}_1,...,\vec{v}_{r}))=r-1
\label{lemma2dimgleichung}
\end{equation}
where $\mbox{Lin}(\vec{v}_1,...,\vec{v}_{r})$ denotes the linear subspace 
spanned by $\vec{v}_1,...,\vec{v}_{r}.$
\\
\emph{Proof.} We can split $\vec{v}_r$ in two parts:
\begin{equation}
\vec{v}_r=\vec{v}_{r\|}+\vec{v}_{r_\bot,}
\end{equation}
where $\vec{v}_{r\|} \in \mbox{Lin}(\vec{v}_1,...,\vec{v}_{r-1})$ and 
$\vec{v}_{r\bot} \bot \; \mbox{Lin}(\vec{v}_1,...,\vec{v}_{r-1}).$ Since
$\vec{v}_r$ is unique, it follows that $\vec{v}_{r\bot}=0$ (otherwise 
$\vec{v}_r=\vec{v}_{r\|}-\vec{v}_{r\bot}$ would be a different solution)
and the first part of the statement is proven. The equality in 
(\ref{lemma2dimgleichung}) comes from the fact that $\vec{v}_{r-1}$ 
is not unique. $\hfill\Box$
\\
\\
{\bf Lemma 3.} Let $M$ be one LvNM in the sense of (\ref{lvnmdefinition})
expanded in the $ G^A_i\otimes G^B_j$ basis:
\begin{equation}
M = \sum_{i,j=1}^N c_{ij}\KetBra{A_{i}}\otimes\KetBra{B_{j}}
  = \sum_{i,j=0}^{N^2-1}\mu_{ij} G^A_i\otimes G^B_j
\end{equation}
Then the $(N^2-1)\times (N^2-1)$ submatrix in the right bottom corner of the
$N^2\times N^2$ matrix $(\mu_{ij})$ (called 
$(\mu_{ij})_{red}=(\mu_{ij})_{i,j=1,...N^2-1}$) has the rank $N-1.$
\\
\emph{Proof.} We can write any of the projectors
$\KetBra{A_{i}}$ and  $\KetBra{B_{j}}$ as Bloch vectors $\hat{A}_i$
and $\hat{B}_j$ (resp. $\vec{A}_i$ and $\vec{B}_j$) with the help of 
(\ref{blochdarstellung}). Then we have 
\begin{equation}
(\mu_{ij})_{red}=\sum_{i,j=1}^N c_{ij} (\vec{A}_i)^T (\vec{B}_j).
\end{equation}
This is a $(N^2-1) \times (N^2-1)$ matrix, since every 
$(\vec{A}_i)^T \vec{B}_j$ is a $(N^2-1) \times (N^2-1)$ matrix.
The range of this matrix is spanned by the vectors  
$(\vec{A}_i)^T.$ The vectors $(\vec{A}_i)^T$ correspond to the 
vectors $\Ket{A_i},$ and they obey relations of the form 
(\ref{orthogonalitaetsbedingung}). Furthermore, $\vec{A}_N$ is uniquely 
determined by $\vec{A}_1,...,\vec{A}_{N-1},$ since $\Ket{A_N}$ 
is uniquely determined by $\Ket{A_1},...,\Ket{A_{N-1}}.$ Thus, we 
can apply our Lemma 2, and the rank of $(\mu_{ij})_{red}$ 
is $N-1.$ $\hfill\Box$ 
\\
\\
{\bf Theorem 2.} Let $\Ket{\psi} \in \HH_A\otimes\HH_B $ have full 
Schmidt rank $N>1.$ Then a  local measurement of the projector 
$\KetBra{\psi}$ requires at least $N+1$ measurements.
\\
\emph{Proof.} If we look at $\KetBra{\psi}$ in the form 
(\ref{lambdaijgleichung}) the matrix $\lambda_{ij}$ has, according
to Lemma 1, the full rank $N^2,$ the reduced matrix 
$(\lambda_{ij})_{red}=(\lambda_{ij})_{i,j=1,...N^2-1}$ has a rank 
of at least $N^2-2.$ 

Since the matrix $(\mu_{ij})_{red}$ corresponding to a single LvNM 
has, according to Lemma 3, the rank $N-1$ we need at least 
$(N^2-2)/(N-1)=N+1-1/(N-1)$ measurements. This proves the statement 
for $N\geq 3.$ For $N=2$ please recall that we have already computed 
that the submatrix $(\lambda_{ij})_{red}$ is 
of rank $N^2-1=3$, not $N^2-2.$ This proves the claim for the 
case $N=2.$ $\hfill\Box$
 
The question remains, which of these results remain valid for 
$N \times M$-systems with $M>N.$ The answer is simple: All results
remain valid. Since the maximal Schmidt rank in a 
$N \times M$-system is $N,$ Theorem 1 can be proven in just the
same way. Also the arguments which led to Theorem 2 can be applied.

\section{Bound entangled states}
In Hilbert spaces with dimensions higher than $2\times 3$, 
there exist entangled states
with positive partial transpose, the {\em bound} entangled states 
\cite{phorod97,horodBE98}.
For this kind of states no general operational entanglement criterion
is known and thus even complete knowledge of the density matrix
may not suffice to decide whether a state is entangled or not.
There exists, however, an important class of bound entangled states,
the so-called "edge" states \cite{edge}, for which the optimal 
witness operators can be constructed explicitly. 
In situations where an experiment is aimed at the generation of an
edge state our method of local decomposition of a witness provides,
therefore, a genuine experimental test.

A state $\delta$ is called an edge
state iff it cannot be represented as $\delta=q\delta'+(1-q)\sigma_{s}$,
where $0\leq q <1$, $\sigma_{s}$ is a separable state and $\delta'$
is a state with a positive partial transpose. In other words, 
for all product vectors $\Ket{e,f}$ and $\epsilon > 0$,
$\delta-\epsilon\KetBra{e,f}$ is not a bound entangled
state anymore. This implies that the edge states lie on the
boundary between the bound entangled states and the
entangled states with non positive partial transpose.
They violate the range criterion \cite{phorod97} in an extremal sense, 
{\em i.e.}
$\delta$ is an entangled edge state with a positive
partial transpose iff for all product vectors 
$\Ket{e,f}\in R(\delta)$,
$ \Ket{e,f^{*}}\notin R(\delta^{T_{B}})$, where $R(\delta)$ denotes
the range of $\delta$.

The generic form of an entanglement witness for such a state 
$\delta$ is \cite{optimization}
\begin{equation}
  \label{eq:ndWitness}
  W=\bar{W}-\epsilon\Eins,
\end{equation}
where
\begin{eqnarray}
  \bar{W}&=&(P+Q^{T_{A}})\\
  \epsilon&=&\inf_{\Ket{e,f}}\Bra{e,f}\bar{W}\Ket{e,f},
\end{eqnarray}
and $P$ and $Q$ denote the projectors onto the kernel of $\delta$
and $ \delta^{T_{A}}$, respectively.

In the following we construct witnesses following this method 
for three kinds of bound entangled edge states  
from the literature, unextendable product basis (UPB) states 
introduced by Bennett {\it et al.} \cite
{Bennett99}, "chessboard" states from Bru\ss \ and Peres 
\cite{Bruss00} and the bound entangled states in $2 \times 4$
dimensions introduced by P. Horodecki \cite{phorod97},
and decompose them locally optimizing the number of projectors
onto product states or the number of settings as explained
above. Note that for this particular construction it is
from an experimentalist's point of view natural to decompose 
and optimize $\bar{W}$ rather than
$W$, because the term $\epsilon\Eins$ does not 
require any special setting.

\subsection{UPB states in $3\times 3$ dimensions}
The states
\begin{eqnarray}
  \Ket{\psi_{0}}&=&\frac{1}{\sqrt{2}}\Ket{0}(\Ket{0}-\Ket{1}), \hspace{0.5cm}
  \Ket{\psi_{2}} = \frac{1}{\sqrt{2}}\Ket{2}(\Ket{1}-\Ket{2}),\nonumber\\
  \Ket{\psi_{1}}&=&\frac{1}{\sqrt{2}}(\Ket{0}-\Ket{1})\Ket{2}, \hspace{0.5cm}
  \Ket{\psi_{3}} = \frac{1}{\sqrt{2}}(\Ket{1}-\Ket{2})\Ket{0},\nonumber\\
  \Ket{\psi_{4}}&=&\frac{1}{3}(\Ket{0}+\Ket{1}+\Ket{2})
  (\Ket{0}+\Ket{1}+\Ket{2})
\end{eqnarray}
form a UPB \cite{Bennett99}, i.e. they are orthogonal to each other and there
is no other product vector orthogonal to all of them.
Therefore the state
\begin{equation}
\varrho_{\textrm{\tiny UPB}}
=\frac{1}{4}(\Eins-\sum_{i=0}^{4}\KetBra{\psi_{i}}),
\end{equation}
which is the projection on the space orthogonal to that
spanned by the UPB, does not contain any product state in
its range. Furthermore, it has a positive partial transpose 
due to the orthonormality of the states $\Ket{\psi_{i}}$.
Therefore, $\varrho_{\textrm{\tiny UPB}}$ is an 
entangled edge state with a positive partial transpose.

The projectors $P$ and $Q$ are related to each other by 
\begin{equation}
 P_{1}=Q_{1}^{T_{A}}=\sum_{i=0}^{4}\KetBra{\psi_{i}},
\end{equation}
therefore we skip $Q^{T_{A}}$ and write the witness as 
\begin{equation}
\label{eq:WUPB}
W_{\textrm{\tiny UPB}}=
\sum_{i=0}^{4}\KetBra{\psi_{i}}-\epsilon\Eins.
\end{equation}
Five measurements are necessary to measure this witness,
one for each of the five projectors, since the UPB is
constructed in such a way that no two projectors can
be evaluated in the same basis.
The main problem of this construction is to find $\epsilon$.
An analytical bound obtained by Terhal \cite{Terhal00} gives
\begin{equation}
 \epsilon\geq\frac{1}{9}\frac{(6-\sqrt{30})}{6}\frac{(2-\sqrt{3})}{2}
  \simeq 0.001297.
  \end{equation}
Numerical analysis leads however to the much bigger value
 $\epsilon\simeq 0.02842$.

In this case it is also interesting to optimize the number
of local projection measurements needed for the measurement 
of the total witness $W_{\textrm{\tiny UPB}}$.
We extend the set $\{\Ket{\psi_{i}},i=0,\ldots,3\}$ with the vectors
\begin{eqnarray}
  \Ket{\psi_{5}}&=&\frac{1}{\sqrt{2}}\Ket{0}(\Ket{0}+\Ket{1}),\hspace{0.5cm}
  \Ket{\psi_{7}} = \frac{1}{\sqrt{2}}\Ket{2}(\Ket{1}+\Ket{2}),\nonumber\\
  \Ket{\psi_{6}}&=&\frac{1}{\sqrt{2}}(\Ket{0}+\Ket{1})\Ket{2},\hspace{0.5cm}
  \Ket{\psi_{8}} = \frac{1}{\sqrt{2}}(\Ket{1}+\Ket{2})\Ket{0},\nonumber\\
  \Ket{\psi_{9}}&=&\Ket{11}
\end{eqnarray} 
to an orthonormal basis which can be used to decompose the identity.
Altogether we are then left with a pseudo-mixture containing
10 projectors. Denoting 
$B_{1}=\{\Ket{0}, \Ket{1}, \Ket{2} \}$,
$B_{2}=\{(\Ket{0}-\Ket{1})/\sqrt{2}, \Ket{2}, (\Ket{0}+\Ket{1})/\sqrt{2} \}$,
$B_{3}=\{(\Ket{1}-\Ket{2})/\sqrt{2}, \Ket{0}, (\Ket{1}+\Ket{2})/\sqrt{2} \}$,
and
$B_{4}=\{(\Ket{0}-\Ket{1})/\sqrt{2}, (\Ket{0}+\Ket{1}+\Ket{2})/\sqrt{3},
(\Ket{0}+\Ket{1}-2\Ket{2})/2\}$,
we easily see that measurement of $W_{\textrm{\tiny UPB}}$ decomposed 
in this form requires 6 correlated settings for Alice and Bob: 
$B_{1}B_{2}, B_{2}B_{1},  B_{1}B_{3}, B_{3}B_{1}, B_{4}B_{4},$ 
and $B_{1}B_{1}$, therefore it is unfeasible to decompose
the whole witness from the number of settings point of view
as noted above.
By subtracting in Eq. (\ref{eq:ndWitness}) some positive
operator $I$ instead of $\Eins$, one can reduce the
number of projectors in the decomposition of $W_{\textrm{\tiny UPB}}$
to 9 -- this gives an ONP, since the number of
terms in any ONP must be larger than or equal to the
rank of the witness, which is equal to 9. The idea
is to form $I$ as a convex sum of projectors onto
$\Ket{\psi_{i}}_{i=0,\ldots,4}$ and onto 4 other
product vectors that are obviously not orthogonal
to the 5 UPB states, but can be chosen such that the 
set of the 9 vectors forms a basis. If we choose
as the additional vectors $\Ket{\bar{\psi}_{i}}_{i=4,\ldots,7}$
the decomposition contains 9 projectors in 5 settings.
The bound for $\epsilon$ has to be adapted as
\begin{equation}
  \epsilon'=\inf_{\Ket{e,f}}\frac{\Bra{e,f}\bar{W}\Ket{e,f}}
  {\Bra{e,f} I\Ket{e,f}}.
\end{equation}
Numerical analysis leads to a value of $\epsilon'\simeq 
0.0311$.
Note that when the bound entangled state is affected
by white noise, namely 
$\rho_{p}=p\cdot\rho_{\textrm{\tiny UPB}}+(1-p)\Eins/9$,
the witness given above is still suitable for the 
detection of entanglement. For the witness in 
Eq. (\ref{eq:WUPB}), $Tr(W_{\textrm{\tiny UPB}}\rho_{p})<0$ when 
$p>(1-9\epsilon/5)$.

\subsection{Chessboard states in $3\times 3$ dimensions}
The states introduced in \cite{Bruss00} are constructed 
from 4 entangled vectors,
\begin{eqnarray}
 \rho_{\textrm{cb}}&=&N\sum_{i=1}^{4}\KetBra{V_{i}},\\
 \Ket{V_{1}}&=&(m,0,s;0,n,0;0,0,0)\nonumber\\
 \Ket{V_{2}}&=&(0,a,0;b,0,c;0,0,0)\nonumber\\
 \Ket{V_{3}}&=&(n^{*},0,0;0,-m^{*},0;t,0,0)\\
 \Ket{V_{4}}&=&(0,b^{*},0;0,-a^{*},0;0,d,0),\nonumber
\end{eqnarray}
where $*$ denotes complex conjugation and 
$N=1/\sum_{j}\langle V_{j}|V_{j} \rangle$. By choosing
the phases of the $\Ket{V_{i}}$ and the basis vectors, 
6 of the parameters can be made real. 
Without loss of generality, $t$ and $s$ can be assumed to be 
complex. In matrix form, $\rho_{\textrm{cb}}$ can then be written 
as
\begin{equation}
\varrho_{\textrm{cb}}=N
  \left(
   \begin{array}{ccccccccc}
     m^{2}+n^{2}&0&ms^{*}&0&0&0&nt^{*}&0&0\\
     0&a^{2}+b^{2}&0&0&0&ac&0&bd&0\\    
     sm&0&|s|^{2}&0&sn&0&0&0&0\\    
     0&0&0&a^{2}+b^{2}&0&bc&0&-ad&0\\    
     0&0&ns^{*}&0&m^{2}+n^{2}&0&-mt^{*}&0&0\\    
     0&ac&0&cb&0&c^{2}&0&0&0\\    
     tn&0&0&0&-tm&0&|t|^{2}&0&0\\    
     0&bd&0&-da&0&0&0&d^{2}&0\\    
     0&0&0&0&0&0&0&0&0
   \end{array}
  \right).
\end{equation}
Bru\ss \ and Peres suggested two methods of ensuring
that $\rho_{\textrm{cb}}$ is bound entangled. We will employ
the first one, i.e. we demand that 
$\rho_{\textrm{cb}}=\rho_{\textrm{cb}}^{T_{A}}$, which
is fulfilled for $t=ad/m$ and $s=ac/n$ real and
implies that $P=Q$.
The kernel of $\rho_{\textrm{cb}}$ (and of $\rho_{\textrm{cb}}^{T_{A}}$) 
is spanned by the (non-normalized) vectors
\begin{eqnarray}
  \Ket{k_{1}}&=&\Ket{22}\\
  \Ket{k_{2}}
     &=&(\frac{m}{n},0,-\frac{m^{2}+n^{2}}{ac};0,1,0;0,0,0)\\   
  \Ket{k_{3}}
     &=&(0,-\frac{ac}{a^{2}+b^{2}},0;-\frac{bc}{a^{2}+b^{2}},0,1;0,0,0)\\   
  \Ket{k_{4}}
     &=&(-\frac{ad}{mn},0,\frac{d}{c};0,0,0;1,0,0)\\   
  \Ket{k_{5}}
     &=&(0,-\frac{bd}{a^{2}+b^{2}},0;\frac{ad}{a^{2}+b^{2}},0,0;0,1,0).   
\end{eqnarray}
In order to decompose the witness $W$ in terms of projectors
onto product states we first examine whether there are more
product vectors in the kernel of $\rho_{\textrm{cb}}$,
so we try to solve 
\begin{equation}
\label{eq:prodeq}
 (a_{1}\Ket{0}+a_{2}\Ket{1}+a_{3}\Ket{2})\otimes(b_{1}\Ket{0}
 +b_{2}\Ket{1}+b_{3}\Ket{2})
 =\sum_{i=1}^{5}x_{i}\Ket{k_{i}}.
\end{equation}
When writing down the equations one can see that the
$x_{i}$ can be substituted by products of one $a_{j}$ 
and one $b_{k}$. Then it is possible to solve the 
set of equations which is then linear in the 
parameters $b_{k}$. This in turn gives two equations for 
the $a_{j}$. The first solution is given by
\begin{equation}
 \Ket{k_{4}'}=\Ket{k_{4}}-\frac{mn}{ac}\Ket{k_{1}}
 =(-\frac{ad}{mn},0,1)\otimes(1,0,-\frac{mn}{ac}),
\end{equation}
and with
\begin{eqnarray}
 \alpha_{1}&\equiv&(m^{2}+n^{2})bmn-(a^{2}+b^{2})am^{2}\nonumber\\
 \alpha_{3}&\equiv&ad^{2}n^{2}\nonumber\\
 \alpha_{13}&\equiv&(m^{2}+n^{2})(mn+ab)d-2abdm^{2}\\
 \gamma_{1}^{0,1}&\equiv&\Big(-\alpha_{13}
   \pm\sqrt{\alpha_{13}^{2}-4\alpha_{1}\alpha_{3}}\Big)/2\alpha_{3}\nonumber\\
 \gamma_{2}^{0,1}&\equiv&\Big[\frac{1}{am^{2}}
 \Big(bmn+d(mn+ab)\gamma_{1}^{0,1}
      +ad^{2}(\gamma_{1}^{0,1})^{2}\Big)
 \Big]^{\frac{1}{2}}
\end{eqnarray}
the other solutions can be written as
\begin{eqnarray}
  \Ket{e,f}=a_{1}\Big(1,\pm\gamma_{2}^{0,1},\gamma_{1}^{0,1}\Big)
    \otimes b_{2}\Big(\pm\frac{m^{2}\gamma_{2}^{0,1}}{mn+ad\gamma_{1}^{0,1}},
    1,\mp\frac{a^{2}+b^{2}+bd\gamma_{1}^{0,1}}{ac\gamma_{2}^{0,1}}\Big).
\end{eqnarray}
The parameters $a_{1}$ and $b_{2}$ can be used to
normalize the vectors. We found 6 product vectors
in the kernels. Of those vectors, 5 will be linearly 
independent in general. Since they do not form
an orthonormal set, we cannot construct $P$ and $Q$
from them. However, the witness can also be constructed
by using instead of the projector onto the kernel
of $\rho_{\textrm{cb}}$ ($\rho_{\textrm{cb}}^{T_{A}}$)
an operator $\tilde{P}$ ($\tilde{Q}$) which is 
strictly positive on the range of the kernel of 
$\rho_{\textrm{cb}}$ ($\rho_{\textrm{cb}}^{T_{A}}$).
This will only affect the value of $\epsilon$.
Here $\tilde{P}=\tilde{Q}$ can be constructed by summing
the projectors onto five linearly independent
product vectors from the kernel of $\rho_{\textrm{cb}}$.
Since the vectors are real, we have in addition that
$\tilde{P}=\tilde{Q}^{T_{A}}$.
Hence in general the "pre-witness" $\bar{W}$ can
be decomposed into 5 projectors onto product vectors
requiring 5 settings to measure $Tr(W_{\textrm{cb}}\rho)$, 
one for each of the projectors.

\subsection{Horodecki states in $2\times 4$ dimensions}
The positive operators introduced in \cite{phorod97} can be written
in matrix form as
\begin{equation}
  \rho_{b}=\frac{1}{7b+1}
  \left(
  \begin{array}{cccccccc}
     b&0&0&0&0&b&0&0\\
     0&b&0&0&0&0&b&0\\
     0&0&b&0&0&0&0&b\\
     0&0&0&b&0&0&0&0\\
     0&0&0&0&\frac{1}{2}(1+b)&0&0&\frac{1}{2}\sqrt{1-b^{2}}\\
     b&0&0&0&0&b&0&0\\
     0&b&0&0&0&0&b&0\\
     0&0&b&0&\frac{1}{2}\sqrt{1-b^{2}}&0&0&\frac{1}{2}(1+b)\\
  \end{array}
  \right),
\end{equation}
where $b\in[0,1]$.
For $b=0,1$, the matrix $\rho_{b}$ is separable, and bound
entangled for all other values of $b$.
In the following we assume $b\neq0,1$.
The kernel of $\rho_{b}$ is spanned by the entangled vectors
\begin{eqnarray}
  \Ket{k_{1}}=\frac{1}{\sqrt{2}}(1,0,0,0;0,-1,0,0)\\
  \Ket{k_{2}}=\frac{1}{\sqrt{2}}(0,1,0,0;0,0,-1,0)\\
  \Ket{k_{3}}=\frac{1}{\sqrt{2+y^{2}}}(0,0,1,0;y,0,0,-1),
\end{eqnarray}
where $y=\sqrt{(1-b)/(1+b)}$, so any vector in the kernel
can be represented as 
\begin{equation}
\Ket{k}=(A,B,C,0;yC,-A,-B,-C)
\end{equation}
where $A,B,C,D$ are complex parameters. For $\Ket{k}$ to
be a product vector, it must be of the form
\begin{equation}
  \Ket{e,f}=(r,s)\otimes(A',B',C',D')\equiv(r(A',B',C',D');s(A',B',C',D')),
\end{equation}
where $r,s,A',B',C'$ are complex parameters. It can be readily
checked that there is no possibility to write $\Ket{k}$ in this
form, therefore there is no product vector in the kernel of $\rho_{b}$.
On the other hand, the kernel of $\rho_{b}^{T_{B}}$ is spanned 
by the entangled vectors
\begin{eqnarray}
  \Ket{k_{1}}=\frac{1}{\sqrt{2}}(0,0,1,0;0,-1,0,0)\\
  \Ket{k_{2}}=\frac{1}{\sqrt{2}}(0,0,0,1;0,0,-1,0)\\
  \Ket{k_{3}}=\frac{1}{\sqrt{2+y^{2}}}(0,1,0,0;1,0,0,y),
\end{eqnarray}
and does not contain any product vector, either.
Therefore, the decomposition of the witness from 
Eq. (\ref{eq:ndWitness}) in projectors onto product vectors
is a rather tedious task. On the other hand, 
we can write down the witness $W$ as in Eq. (\ref{eq:ndWitness}) and then 
decompose it as in Eq. (\ref{tensorproduktzerlegung}) and
(\ref{einfachezerlegung})
\begin{equation}
  W=\sum_{i=0}^{3}\sum_{j=0}^{15}w_{ij}\sigma_{i}\otimes\tau_{j}
   \equiv\sum_{i=0}^{3}\sigma_{i}\otimes\tilde{\tau}_{i}
\end{equation}
in a straightforward manner. Here $\sigma_{i}$ and $\tau_{i}$,
for $i>0$, are the generators of the SU(2) and SU(4), respectively,
while $\sigma_{0}=\Eins_{2}/2$ and $\tau_{0}=\Eins_{4}/4$. 

The matrices $\tilde{\tau}$ turn out to be
\begin{equation}
  \tilde{\tau}_{0}=
    \left(
    \begin{array}{cccc}
     -c&0&0&0\\
     0&c&0&0\\
     0&0&c&0\\
     0&0&0&-c
    \end{array}
    \right)
    +(6-8\epsilon)\tau_{0},\\
\end{equation}

\begin{equation}
  \tilde{\tau}_{1}=
  \frac{1}{4}
  \left(
  \begin{array}{cccc}
    0&-cy^{2}&2cy&0\\
    -cy^{2}&0&-2&2cy\\
    2cy&-2&0&-c(4+y^{2})\\
    0&2cy&-c(4+y^{2})&0
  \end{array}
  \right),
\end{equation}

\begin{equation}
  \tilde{\tau}_{2}=\frac{i}{4}
    \left(
      \begin{array}{cccc}
        0&-cy^{2}&-2cy&0\\
        cy{2}&0&-2&-2cy\\
        2cy&2&0&-c(4+y^{2})\\
        0&2cy&c(4+y^{2})&0
      \end{array}
   \right),
\end{equation}
and
\begin{equation}
  \tilde{\tau}_{3}=-cy^{2}\tau_{0},
\end{equation}
where $c=1/(2+y^{2})$.
The number of correlated 
settings is 4, but Alice and Bob need  
only 3 different settings each to measure the witness.

Summarizing, we have shown that for UPB states
and "chessboard" states in $3\times 3$ systems,
one needs 5 LvNMs to detect the witness, whereas
for Horodecki states in $2\times 4$ systems,
4 correlated measurements are necessary.

\section{Conclusion}
In this paper we have studied the problem of detection of 
entangled states using entanglement witnesses and few local 
measurements. Two optimization scenarios were discussed:
the one corresponding to the decomposition of the witness operator 
into the optimal number of projectors on product vectors (ONP)
and the one corresponding to the decomposition of the witness 
into the optimal number of settings of detecting devices (ONS). 
Several exact results and
estimates have been obtained concerning optimal detection 
strategies for NPT states on $2 \times 2$ and $N \times M$ systems, 
entangled states in 3 qubit systems, and bound entangled states in 
$3\times3$ and $2\times4$ systems. Despite numerous results and 
progress in understanding this problem, the general question of finding
ONP and ONS for an arbitrary witness operator remains open.

We wish to thank K. \.Zyczkowski,
I. Cirac, S. Haroche, S. Huelga, B. Kraus, 
and H. Weinfurter for  discussions. This work has been supported  
by  DFG (Graduiertenkolleg 282 and Schwer\-punkt  
``Quanteninformationsverarbeitung"), 
the ESF-Programme PESC,  and the EU IST-Programme
EQUIP.

While finishing this paper we became aware of a recent preprint
by A. Pittenger and M. Rubin \cite{piru}, where it was shown how, 
if $N$ is prime, a projector onto a pure maximally entangled state 
with full Schmidt 
rank can be measured with $N+1$ LvNMs. Thus, our bound from 
Theorem 2 can be reached for this case.

\end{document}